\documentclass[twoside]{dis09}
\usepackage[latin1]{inputenc}
\usepackage[dvips]{graphicx,epsfig,color}
\usepackage{wrapfig,rotating}
\usepackage{amssymb,amsmath,array}

\begin{document}
\title{Physics Program with Tagged Forward Protons \\
 at STAR/RHIC}

\author{J.H. Lee for the STAR Collaboration
%
\thanks{This work was supported by Brookhaven Science Associates, LLC under Contract
No. DE-AC02-98CH10886 with the U.S. Department of Energy.}
%
\vspace{.3cm}\\
%
Brookhaven National Laboratory - Physics Department \\
Upton, NY 11973 - U.S.A.
}

\maketitle

\begin{abstract}
A new effort to explore the diffractive regime in polarized $p+p$ collisions in a broad high energy range 
($\sqrt{s}$ = 200 - 500 GeV) has been initiated with the STAR detector at RHIC.
Staged implementation of multiple Roman Pot stations for tagging the forward proton in the diffractive processes
will enable searches for the centrally produced for the possible gluon bound state via double 
Pomeron exchange process and the theoretically expected Odderon state in QCD 
by studying  spin-dependent elastic scattering in a wide $t$-range with polarized $p+p$.  
\end{abstract}

\section{Introduction}
Diffractive processes at high energies are believed to be    
occurring via the exchange of a color singlet object (the ``Pomeron'')  with the same
quantum numbers as the vacuum ($J^{PC}$ = $0^{++}$)~\cite{pomeron}. The Pomeron is considered as a dynamical system 
rather than a particle which is expected to dominate exchange mechanisms at asymptotic energies. 
Since there is no color exchanged between the 
Pomeron and the parent nucleon, a rapidity gap in the final state emerges as a characteristic signature 
of diffractive events. Depending on the distribution of the gap and number of outgoing particles with same quantum number 
of initial particles, diffractive processes can be classified as elastic scattering, single diffraction (SD), 
or double diffraction (DD). 
Even though properties of diffractive scattering are described by the phenomenology of Pomeron exchange 
in the context of Regge theory, the exact nature of the Pomeron still remains elusive. 
Main theoretical difficulties in applying QCD in diffraction are due to the intrinsically non-perturbative nature of the process 
in the kinematic and energy ranges of the data currently available.
The experimental challenge is identifying and reconstructing forward protons kinematically very close to the beam. 

The diffractive physics program at the Relativistic Heavy-Ion Collider (RHIC) is a new experimental program to study 
elastic and inelastic diffractive processes (SD, DD) with tagged forward protons in polarized  
$p+p$ collisions at $\sqrt{s}$ = 200 - 500 GeV.  
For the elastic program, the collider energy range is previously unexplored, and the measurements will serve 
as an important bridge between vast lower energy data and limited measurements at higher energy data. 
The energy range, particularly with polarized $p+p$ collisions, is suitable as a testing ground for the long standing 
theoretical evidence in QCD for the existence of the Odderon which is the $C=P=-1$ counterpart to the Pomeron~\cite{pomeron}.   
The main physics motivation 
for the inelastic diffraction program is for searching for a gluonic bound state whose existence is allowed in pure gauge 
QCD, but for which  no unambiguous candidate has been established.

\begin{wrapfigure}{r}{0.5\columnwidth}
\centerline{\includegraphics[width=0.47\columnwidth]{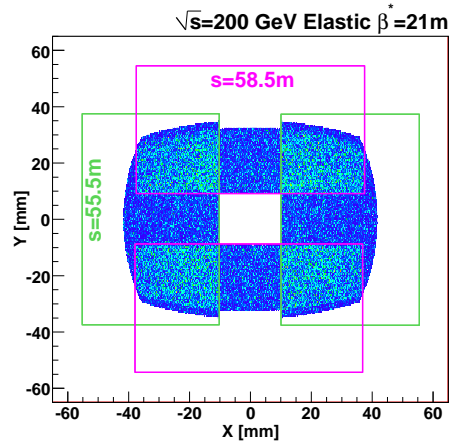}}
\caption{Y vs. X  distribution of protons in the RPs for elastic events for the Phase I setup, 
which was used for RHIC/Run-9.}\label{Fig:xy}
\end{wrapfigure}
\section{Tagging Forward Protons in Diffractive Processes}
 Although identification of a rapidity gap can be utilized for studying diffractive processes, 
it is imperative to tag and reconstruct the  forward proton to eliminate the ambiguity of a rapidity gap tag, 
which can be contaminated by background due to low multiplicity non-diffractive processes. 
The rapidity gap tag also does not provide information on whether the initial proton remains intact after the collision 
or is excited into a low-mass state with small energy loss, which could still yield a rapidity gap.
Tagging forward protons, i.e., detecting scattered protons in diffractive process
requires reaching inside of the beam pipe since the scattering angles are very small 
of the order of $\mu$rad to a few mrad.  This can be  achieved by exploiting the technique 
of the ``Roman Pot''(RP) which has been 
developed and used at the ISR~\cite{amaldi}. At RHIC, the technique has been successfully used for 
studying elastic scattering in the experiment pp2pp~\cite{pp2pp}.
By detecting the scattered proton in RPs, one can reconstruct its momentum from 
the measured  positional and directional information of the protons in the given beam optics and 
thus derive two key variables characterizing kinematics of the collisions:  
four momentum transfer $t = (P_1 - P)^2$ and 
longitudinal momentum fraction $\xi=(1-x_F)$, where $x_F = 2p_L/\sqrt{s}$. 
$\xi$ is simply related to the  momentum fraction of the proton carried by the Pomeron 
(${I\!\!P}$), commonly refereed as $x_{{I\!\!P}}$.

\begin{wrapfigure}{r}{0.5\columnwidth}
\centerline{\includegraphics[width=0.47\columnwidth]{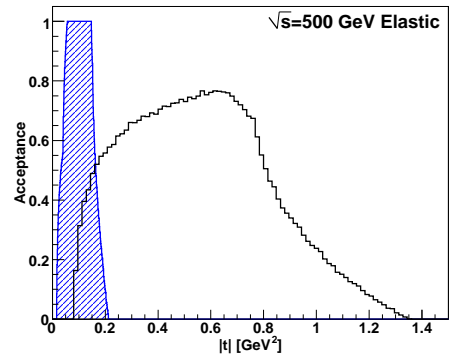}}
\caption{Acceptance of elastic events as a function of $t$ for Phase I (hatched) and Phase II at $\sqrt{s}$=500 GeV.}\label{Fig:acc}
\end{wrapfigure}
For a wide kinematic coverage, a staged implementation of RP is considered.    
For the initial phase of the new program (Phase I), which probes for probing small-$t$ reaction, 
the Roman Pots used for the pp2pp experiment have been  
integrated with STAR detector~\cite{star}.   
They are positioned  at $\pm 55.5$, $\pm 58.5$m from the nominal interaction point (IP). Each
RP contains four planes of silicon strip detectors (SSD)  (two vertical and two horizontal)
to provide redundancy for the track reconstruction. Figure~\ref{Fig:xy} shows the distribution of accepted positions
of the elastically scattered protons in SSD in two RP stations at  $\sqrt{s}$ = 200 GeV with the beam $\beta^{*}$=21m 
for minimum angular divergence.  
The corresponding accepted $t$ range is 0.002 $<|t|<$ 0.03 GeV$^2$.
During RHIC Run-9, $\sim$70M  of events, including  $\sim$30M of elastic events, 
were successfully taken with the Phase I set-up, and  the data are currently being analyzed.

\begin{wrapfigure}{r}{0.6\columnwidth}
\centerline{\includegraphics[width=0.55\columnwidth]{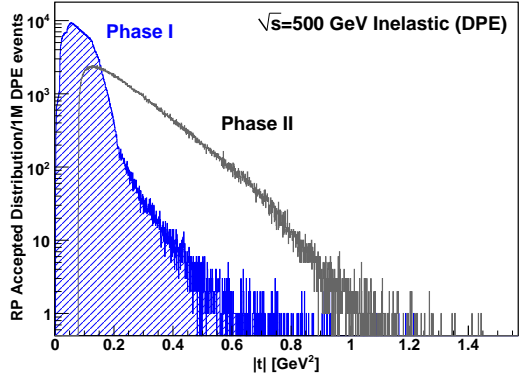}}
\caption{Accepted $t$-distributions for inelastic diffractive collisions at $\sqrt{s}$=500 GeV with Phase I 
and Phase II setups.}\label{Fig:acc_t}
\end{wrapfigure}
For Phase II, new sets of  Roman Pot system will 
be designed and fabricated to be  installed between RHIC DX-D0 magnets, 
15-17m from the IP, extending  the acceptance and the reach in $t$ and  $\xi$ 
for a more optimized setting for inelastic diffractive program.  
Figure~\ref{Fig:acc} shows the acceptance for 
elastic events for the Phase I and Phase II Roman Pot setups, and Fig.~\ref{Fig:acc_t} 
shows the accepted $t$-distributions for  
inelastic diffractive events with the Phase I and Phase II Roman Pot setups.  

For the inelastic diffractive studies, the Roman Pot systems will be used in conjunction with 
the STAR TPC to reconstruct and fully
constrain events with a resonance in central production process.  Since no special accelerator optics
is required in this configuration for the Phase II set-up, running in parallel with other physics program in 
STAR is possible, and we will be able to utilize high luminosity for searches for rare physics processes.
\section{Elastic Scattering: Searching for the Odderon} 
In elastic collisions with very small $t$, there is  interference
between hadronic spin-flip helicity amplitudes, $\phi_2=\langle +\!+|T|-\!-\rangle$ and  
$\phi_4 = \langle +\!-|T|-\!+ \rangle$, and  the electromagnetic non-flip amplitude.
This provides a sensitive tool to study the spin dependence of diffractive scattering
at asymptotic energies and to search of the  Odderon exchange~\cite{trueman}. 
The Odderon is a $CP$ odd counterpart of the Pomeron which has been postulated from theoretical considerations, 
but has not yet been established experimentally.
Because Pomeron and Odderon have opposite $C$-parities, it is expected in leading order, that if
Pomeron and Odderon have the same asymptotic behavior, they are out
of phase by approximately 90$^\circ$~\cite{leader,buttimore}. Therefore, if they couple to spin, their interference with
the electromagnetic non-flip amplitude will result in different $t$-dependences of the double
spin asymmetries, $A_{NN}$~\cite{trueman}.   
The data taken with Phase I setup during the RHIC run-9 are expected to produce precise measurements of 
$A_{NN}$ distributions covering the region where the strong $t$-dependence of the asymmetry is predicted.      
It is also predicted that there will be a difference in total cross-section and $d\sigma/dt$ in $p+p$ and $\bar{p}+p$ 
collision in $0<|t|<1.5$ GeV$^2$ 
if the Odderon exists.  The measurements~\cite{us4} by UA4 in $\bar{p}+p$ at $\sqrt{s}=546$ GeV 
can be compared with the RHIC data in $p+p$ at $\sqrt{s}$ = 500 GeV $\bar{p}+p$. Even though there will be 
$\approx$10$\%$ of difference ($\Delta  \sigma_{tot}$) between the two energies, the overlap between the two data sets
will make the measurements an unique and exciting opportunity for this challenging study~\cite{islam}.
Since the predicted $\Delta  \sigma_{tot}$ is small ($\approx$3 mb) and the most notable difference in $d\sigma/dt$
will be at  $-t$$\approx$1 GeV$^2$, the measurements require accurate measurement in a wide $t$ range requiring both Phase I and Phase II setups.   
\section{Inelastic Diffractive Double Pomeron Exchange Process: Searching for the Glueball} 
In the context of QCD, the Pomeron exchange is believed to be the exchange of a system of gluons. 
The double Pomeron exchange (DPE) process has been
regarded as one of the potential channels of glueball production~\cite{glueball}.   
Two of the gluons in the DPE process could merge into a mesonic bound 
state without a constituent quark, a glueball in the central production $p+p\rightarrow p+M_{X}+p$.
The central rapidity is expected to be spanning ln$M_{x}$ and the rapidity gap in DPE process  
is $y_{beam}$$-$$y_{central}$$\approx$3 units, which allows the maximum kinematical value of $M_X$$\approx$25 GeV/$c^2$ with
clean rapidity gaps.

Lattice QCD calculations  have predicted the lowest-lying scalar glueball state in the mass range of 
1500-1700 MeV/$c^2$, and tensor and pseudoscalar glueballs in  2000-2500 MeV/$c^2$~\cite{glueball}. 
Experimentally measured glueball candidates for the scalar glueball states are the
$f_0(1500)$ and the $f_J(1710)$~\cite{abatziz}  in central production as well as other gluon-rich reactions
such as $\bar{p}p$ annihilation, and radiative $J/\psi$ decay~\cite{glueball_exp}. 
The spin of the $f_J(1710)$ is not yet confirmed, indications for both spin 2 and spin
0 have been reported. 
The glueballs are expected to be intrinsically unstable and decay in diverse ways, yielding typically two or more mesons. 
$f_J(1710)$ dominantly decays into $K^+K^-$ and $f_0(1500)$ into $\pi^+\pi^-\pi^+\pi^-$.
\begin{figure}{}
\centerline{\includegraphics[width=0.7\columnwidth]{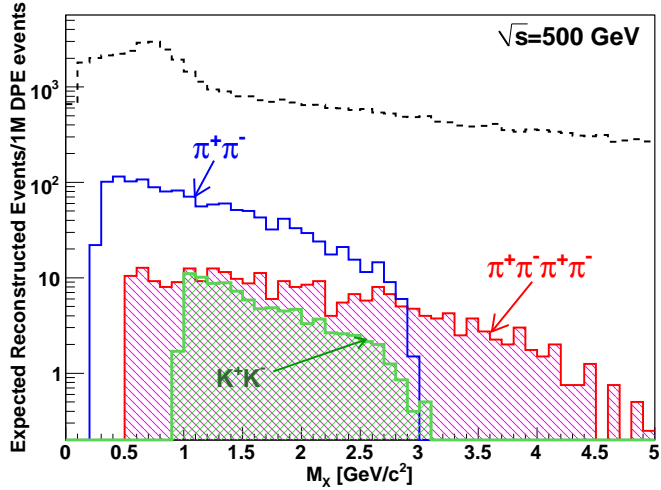}}
\caption{Estimated phase-space distributions in effective mass for $M_X$ decaying to 
$\pi^+\pi^-$, $\pi^+\pi^-\pi^+\pi^-$ (hatched), and $K^+K^-$ (cross-hatched) from 1M  DPE events in $p+p\rightarrow p+M_{X}+p$ at 
$\sqrt{s}$ = 500 GeV. The dotted line shows the effect mass distributions generated from the DPE events accepted 
in the Roman Pots (Phase II). }\label{Fig:eff}
\end{figure}
One of the challenges in identifying a glueball state unambiguously lies in difficulties of
isolating a glueball state from the conventional meson state that shares the same quantum numbers. 
To identify that the process is from DPE process rather than Reggion exchange requires observing 
suppression of $\rho$ meson in the process, since $\rho$ cannot be formed from two
states with $J^{PC}=0^{++}$. The other ``filter'' for enhancing glueball candidates in DPE process is
the ``$\delta P_T$'' filter~\cite{dptfilter}, in which small momentum transfer processes enhance $gg$ 
kinematic configurations since the gluons can flow directly into the final state in the process.  
The technique for reconstructing resonances using the STAR detector system has been well established in $p+p$ and $A+A$ collisions.
Especially the current ongoing central photo-production program in ultra-peripheral $AA$ collisions at STAR~\cite{star_upc} 
is topologically similar to the DPE process, and the common experimental machinery can be utilized for triggering and
analyzing DPE processes.   
 
Figure~\ref{Fig:eff} shows the expected reconstructed kinematic phase-space distributions of centrally produced 
mass decaying into $\pi^{+}\pi^{-}$ for 1M DPE events. Reconstruction of the tracks from decay was simulated using
geometrical acceptance of the TPC ($-1<y<1$ with full azimuthal coverage) and particle identification by TPC and the 
Time-of-Flight system which can separate $\pi/K$ in momentum range up to 1.6 GeV$/c$.
The effective mass range 1-3 GeV$/c^2$ are kinematically well accessible in pion and kaon decay channels.
The high mass region is limited by particle identification and particles decaying outside of the rapidity coverage of TPC ($-1<y<1$).
Expected trigger rates for DPE are 80 Hz at $1\times10^{31}$cm$^{-2}$s$^{-1}$. The two-Pomeron (${I\!\!P}$${I\!\!P}$) cross-section 
at the RHIC energy is not known and  estimates in Ref.~\cite{streng,simonov} are used.  
To collect $\sim$100K $K^+K^-$ and $\sim$250K $\pi^+\pi^-\pi^+\pi^-$  data sample ($\sim$75K, $\sim$100K in 1$<M_X<2$ GeV/$c^2$, respectively), 
it is estimated to require 2-3 weeks of RHIC running time reassuming branching ratios of DPE processes measured at 
$\sqrt{s}$ = 62.4 GeV~\cite{branching}. The assumed integrated luminosity can be easily achieved during the planned  
high luminosity spin program at RHIC in the future, and it's expected that the luminosity upgrade and longer run can bring an order of 
magnitude higher statistics which will enable differential kinematic sampling and spin-parity analysis.
The Odderon state can also be identified by studying centrally produced $C=-1$ diffractive vector mesons such as
$J/\psi$ produced by Odderon-Pomeron coupling mixed with the photon-Pomeron coupling in inelastic diffractive 
collisions~\cite{bzdak}. The measurement is expected to be deliverable by the 
STAR detector upgrade for optimized measurements for $J/\psi$ and high luminosity running  
even though the expected cross-section for diffractive $J/\psi$ at the RHIC energy is small~\cite{klein}. 
\section{Summary}
A new rich diffractive physics program in polarized $p+p$ at RHIC with tagged forward protons
using the Roman Pot technique with the STAR detector system has been launched.
The unique machine and detector capabilities enable us to explore an important aspect 
of our understanding of the strong interaction. The main physics motivation is to search for theoretically predicted
states in QCD: the Odderon and the glueball. The diffractive program, together with the other RHIC physics programs, 
will serve as an important role toward our complete understanding of the strong interaction
and quantum chromodynamical description of the hadronic structure.
 

\begin{footnotesize}


%

\end{footnotesize}


\end{document}